\documentclass[twocolumn,aps,superscriptaddress,multicol,amsmath,amssymb]{revtex4-1}
\usepackage{dsfont,amsthm,amsbsy}
\usepackage{amssymb}
\usepackage{amsmath}
\usepackage{graphicx}
\usepackage{epstopdf}
\usepackage{subfigure}
\usepackage{epsfig}
\usepackage{amsfonts}
\usepackage{mathrsfs}
\usepackage{sidecap}
\usepackage{lipsum}
\usepackage[colorlinks,linkcolor=blue,citecolor=blue,anchorcolor=blue]{hyperref}
\usepackage{dsfont,amsthm,amsbsy}
\usepackage{resizegather}
\usepackage{tikz}
\usepackage{float}

\newcommand{\bl}{\begin{aligned}}
\newcommand{\el}{\end{aligned}}
\def\be{\begin{equation}}
\def\ee{\end{equation}}

\def\bi{\begin{itemize}}
\def\ei{\end{itemize}}
\def\bn{\begin{enumerate}}
\def\en{\end{enumerate}}
\def\bea{\begin{eqnarray}}
\def\eea{\end{eqnarray}}
\def\no{\nonumber}
\def\ba{\begin{array}}
\def\ea{\end{array}}
\def\bd{\begin{displaymath}}
\def\ed{\end{displaymath}}

\usepackage{hyperref,cleveref}

\graphicspath{{FigsN/}} 

\begin{document}
\title{Geometrically Frustrated Anisotropic Four-Leg Spin-1/2 Nanotube}
\author{R. Jafari}
\email[]{jafari@iasbs.ac.ir}
\email[]{rohollah.jafari@gmail.com}
\affiliation{Department of Physics, Institute for Advanced Studies in Basic Sciences (IASBS), Zanjan 45137-66731, Iran}
\affiliation{Department of Physics, University of Gothenburg, SE 412 96 Gothenburg, Sweden}
\affiliation{Beijing Computational Science Research Center, Beijing 100094, China}
\author{Saeed Mahdavifar}
\affiliation{Department of Physics, University of Guilan, 41335-1914, Rasht, Iran}
\author{Alireza Akbari}
\affiliation{Asia Pacific Center for Theoretical Physics (APCTP), Pohang, Gyeongbuk, 790-784, Korea}
\affiliation{Department of Physics, POSTECH, Pohang, Gyeongbuk 790-784, Korea}
\affiliation{Max Planck POSTECH Center for Complex Phase Materials, POSTECH, Pohang 790-784, Korea}
\affiliation{Department of Physics, Institute for Advanced Studies in Basic Sciences (IASBS), Zanjan 45137-66731, Iran}

\date{\today}
%
\begin{abstract}
We develop a   real space quantum  renormalization group (QRG)  to explore a  frustrated anisotropic four-leg spin-1/2 nanotube in the thermodynamic limit.
We obtain the phase diagram, fixed points, critical points, the scaling of coupling constants and magnetization curves.
Our investigation points out that in the case of strong leg coupling the diagonal frustrating interaction is marginal under QRG transformations and does not affect the universality class of the model.
Remarkably, the renormalization equations express that the spin nanotube prepared in the strong leg coupling case goes to the strong plaquette coupling limit (weakly interacting plaquettes).
Subsequently, in the limit of weakly interacting plaquettes, the model is mapped onto a 1D spin-1/2 XXZ chain in a longitudinal magnetic field under QRG transformation.
Furthermore,   the effective Hamiltonian of the spin nanotube inspires both first and second order phase transitions accompanied by the fractional magnetization plateaus.
Our results show that the anisotropy changes the magnetization curve and the phase transition points, significantly.
Finally, we report the numerical exact diagonalization  results to compare the ground state phase diagram with our  analytical visions.
\end{abstract}
\maketitle

\section{Introduction\label{sec1}}
%
The induces non-trivial magnetic states are  currently renewed interest in  the magnetic  quantum spin systems that exhibit a geometrical frustration~\cite{Book1, Book2}.
These states are fascinating because of their intriguing and unique properties compare to conventional magnetic systems, e.g.,  
 unconventional magnetic orders or even a disorder~\cite{Book1, Book2, Jafari2007}.
They have attracted more attention by experimental realization of $J_{1}-J_{2}$ chain~\cite{Hase},
whereas  a geometric frustrated spin can be advanced to a family of problems concerns the integer-spin ladders~\cite{Cabra1998,Dagotto:1996aa,Dagotto:1999aa}.
A nice   extension of two-leg spin ladders   have been performed for the various quantum $n$-leg spin ladder, which are recognized as  
 tubelike lattice structures  for $n\geq 3$~\cite{Uehara:1996aa,Cabra1998,Charrier:2010aa}.
Therewith, according to the Lieb-Schultz-Mattis theorem, their ground state can be either gapped with a broken translational invariance, or gapless (non-degenerate)~\cite{Lieb,Arlego2013}.
\\

\begin{figure}[t]
\includegraphics[width=1.0\columnwidth]{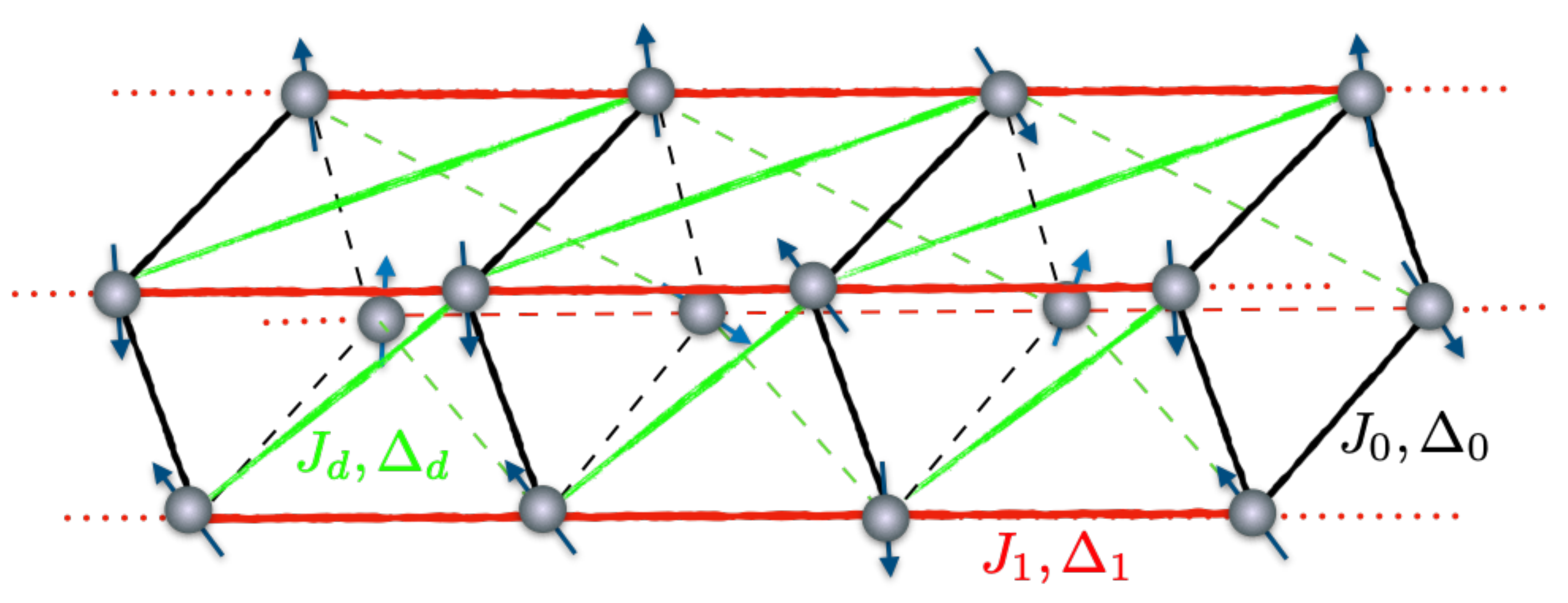}
\caption{(Color online) Schematic representation of the frustrated four-leg spin tube.
The interaction along the legs is characterized by $J_{1}$ (red lines),  $J_{0}$ shows the intra-plaquettes infraction (black lines), and
the diagonal interaction $J_{d}$ has been shown by the green lines.
}
\label{fig1}
\end{figure}

An illustrative example is    a triangular frustrated structure, which   can be found in the  three-leg spin tubes with the larger frustration and quantum fluctuations~\cite{Kawano:1997aa,Wang:2001aa}.
These  three-leg spin tubes have been studied intensively both experimentally and theoretically~\cite{Kawano:1997aa,Wang:2001aa,Sato:2007aa,Luscher:2004aa,Sato:2007ab,Sakai:2008aa,Manaka:2011aa,Hagihala:2019aa,Ochiai:2017aa,Seki:2015aa},
 and have been developed by synthesizing of odd number ($n$) of the legs spin tube, such as [(CuCl$_{2}$tachH)$_{3}$Cl]Cl$_{2}$~\cite{Schnack}, and  XCrF$_{4}$~(X=Cs~or~K) with $n=3$~\cite{Manaka}, as well as,  Na$_{2}$V$_{3}$O$_{7}$
 with $n=9$~\cite{Millet}.
One of the most recent example is a four-leg spin-1/2 nanotube Cu$_{2}$Cl$_{4}$·D$_{8}$C$_{4}$SO$_{2}$ with next-nearest neighbor (NNN) AFM interaction, and
diagonally coupling adjacent legs~(Fig.~\ref{fig1})~\cite{Garlea, Zheludev, Garlea2009}. 
Although,   the spin tubes with an odd number of legs and only nearest neighbor antiferromagnetic (AFM) intrachain coupling are geometrically frustrated,
it is well known that the four-leg spin tube with only nearest neighbor AFM exchange is not frustrated,
neither in the weak or the strong plaquette coupling limits~\cite{Arlego2013,Arlego2011,Cabra1997,Cabra1998,Totsuka1997,Kim1999}.
\\

%

%

%

The frustration in  four-leg spin tube arrises by considering the next nearest coupling or diagonal interaction (see $J_d$, green lines, in Fig.~\ref{fig1}).
This  has generated much recent interest, and  there have been several theoretical attempts to look at the frustrated isotropic four-leg spin tube (FAFST)~\cite{Arlego2011,Arlego2013,Gomez,Rosales:2014aa,Plat:2015aa}. In particular, it  has been investigated using the density
matrix renormalization group (DMRG)\cite{Gomez}, exact diagonalization, series expansion\cite{Arlego2011}, Schwinger bosons mean field theory, and quantum Monte-Carlo simulation~\cite{Arlego2013}.
Moreover, a study of the phases of the FAFST in the presence of a magnetic field has been done in a combined analysis using perturbative methods, variational approach and DMRG~\cite{Gomez}.

Although the magnetic properties of the FAFST model has been investigated in previous works, still the perceptive of the quantum phases on a larger scale is  missing~\cite{Arlego2013}.
In this light, it is inexplicable that  the phase diagram is not fully understood, and also the universality class of the model is unknown in the presence of anisotropy.
This encourages us to  investigate the FAFST model in the presence of a magnetic field using the real space renormalization group (RSRG) approach.   \\

In this respect, we demonstrate
the ground state magnetic phase diagram of such a system, and aim to  show that in the limit of the strong leg coupling
the diagonal frustrating interaction 
 does not flow under QRG transformation.
Furthermore, 
in the limit of the strong leg coupling, the spin tube goes to the strong plaquette coupling limit (weakly interacting plaquettes) under renormalization transformations.
Subsequently, in the strong plaquette coupling limit, under RSRG transformation, the FAFST model maps onto the one-dimensional (1D) spin-1/2 XXZ model in the presence of an effective magnetic field.
We also  show that when the leg and frustrating couplings are the same (maximum frustration line), only first order quantum phase transitions is observed at zero temperature.
 This results that the magnetization (per particle) process exhibits fractional plateaus at zero, one-quarter, one-half and three-quarter of the saturation magnetization.
  We find that away from the maximum frustration line the model exhibits both first and second order quantum phase transitions.
In addition, the numerical Lanczos method is applied for the finite size spin-1/2 nanotubes and mentioned behavior is approved.

%
%
\begin{figure}
\includegraphics[width=1.0\columnwidth]{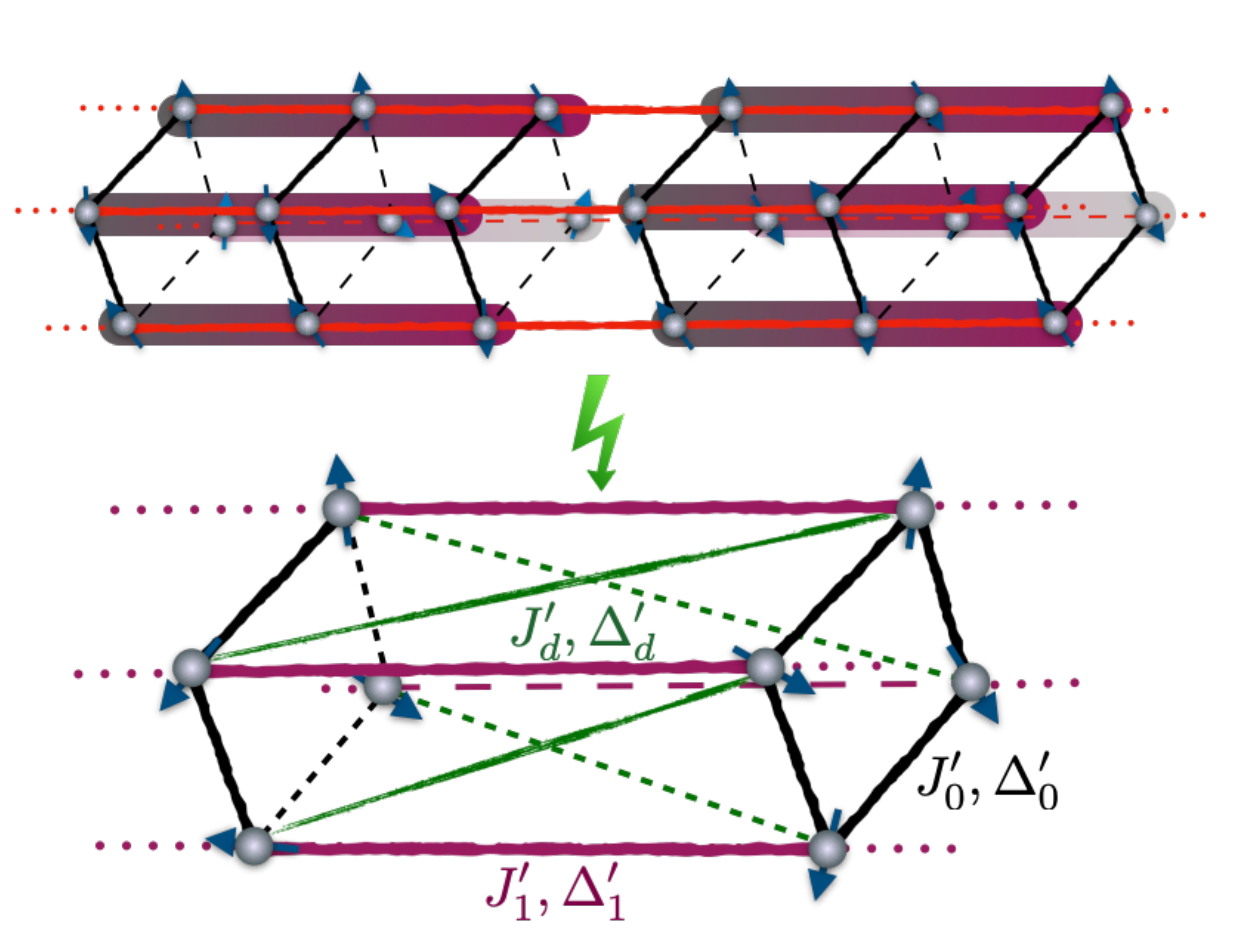}
\caption{(Color online) Renormalization scheme of frustrated four-leg nano-spin tube in
the strong leg-coupling limit, where a three site blocks Hamiltonian along the legs (top) are mapped to a
renormalized spins (bottom).}
\label{fig2}
\end{figure}
%

\section{ FAFST model: Real Space  Renormalization Group Study}

\subsection{Theoretical Model}
We consider the Hamiltonian of the geometrically frustrated anisotropic four-leg spin tube (FAFST) model in the presence of a magnetic
field on a periodic tube of $N$ sites, which is given by
%
\begin{eqnarray}
\bl
\label{eq1}
{\cal H}=
\frac{1}{4}\sum_{i=1}^{N}\sum_{\alpha=1}^{4}
\Big[
&
{\cal H}_{0}^{(i,\alpha;\; i,\alpha+1)}
+
 {\cal H}_{1}^{(i,\alpha;\; i+1,\alpha)}
\\&
+
{\cal H}_{d}^{(i,\alpha;\; i+1,\alpha+1)}
-2h\sigma^{z}_{i,\alpha}
\Big].
 \\
\el
\end{eqnarray}
%
%
Here  we define
%
\begin{equation}
\label{eq2}
{\cal H}_{n}^{(i,\alpha;\; j,\beta)}= J_{n}
(
\sigma^{x}_{i,\alpha}\sigma^{x}_{j,\beta}
+
\sigma^{y}_{i,\alpha}\sigma^{y}_{j,\beta}
)
+
\Delta_{n}
\sigma^{z}_{i,\alpha}\sigma^{z}_{j,\beta},
\end{equation}
with $n=(0,1,\; d)$, the indeses $\alpha, \beta= 1-4 $ run over  intra-plaquettes spins and $i,j$ count the inter-plaquettes sites. Here  $J_{0}$, $J_{1}$ and $J_{d}$ are the plaquette, leg, and diagonal exchange couplings respectively, and the corresponding easy-axis anisotropies are defined by $\Delta_{0}$, $\Delta_{1}$ and $\Delta_{d}$.
Furthermore, $\vec{\boldsymbol \sigma}=(\sigma^x,\sigma^y,\sigma^z)$ denotes the Pauli matrices, and
$h $  represents a magnetic field to point along the $z$-direction.

The FAFST model is invariant under several symmetry operations: one-site translation $\vec{{\boldsymbol \sigma}}_{i,\alpha} \rightarrow \vec{{\boldsymbol \sigma}}_{i+1,\alpha}$, bond-centered inversion $\vec{{\boldsymbol \sigma}}_{i,\alpha} \rightarrow \vec{{\boldsymbol \sigma}}_{1-i,\alpha}$, time reversal $\vec{{\boldsymbol \sigma}}_{i,\alpha} \rightarrow -\vec{{\boldsymbol \sigma}}_{i,\alpha}$, $U(1)$ spin rotation around the $z$ axis, and $\pi$ rotation around the $x$ or $y$ axis~\cite{Fuji:2016aa}.
%
We should mention that the  isotropic form of the above Hamiltonian, $\Delta_{n=0,1,d}=1$, 
has the $SU(2)$ symmetry in the absence of the magnetic field.
%

%
Based on real Space  Renormalization group approach (Appendix~\ref{App_A}), and using  the low-energy effective Hamiltonian we discuss  the properties of the
spin tube,  Eq. (\ref{eq1}) in  two possible limits.
The first limit that  is considered in the section~\ref{Se-strong-limit}, corresponds to the strong leg-coupling  and  is defined by
$J_{1}/J_{0,d}\rightarrow\infty$.
The second case,   is the strong plaquette coupling limit
$J_{0}/J_{1,d}\rightarrow\infty$ and is reviewed in the section~\ref{Se-weak-limit}, which corresponds to almost decoupled plaquettes.
%

\subsection{The strong leg coupling limit: $J_{1}/J_{0,d}\rightarrow\infty$}
\label{Se-strong-limit}
 In this regime the spin nanotube resembles a chain of four weakly
coupled XXZ chains where the weak plaquette coupling $J_{0}$ and diagonal coupling $J_{d}$ can be treated perturbatively. We study this limit for two different cases of in the absence and in the presence of external fields:

\subsubsection{In the absence of the magnetic field: $h=0$}
As a first step, to get a better understanding of the spin tube properties in the strong
leg-coupling limit, we look at the Hamiltonian without the magnetic field, $h=0$.
To implement the idea of QRG in the strong leg-coupling limit, we use the three site block each along every leg (Fig. \ref{fig2}) and kept the degenerate ground states of each block to construct the projection operator~\cite{Jafari2007}.
The inter-block Hamiltonian ${\cal H}^{BB}$, the block Hamiltonian $h^{B}_{I}$ of the three sites and its eigenstates and eigenvalues
are given in Appendix~\ref{App_B}.
Calculating the effective Hamiltonian to the first order correction leads to the effective renormalized Hamiltonian exactly similar to the initial one, Eq.~(\ref{eq1}), by exchanging the couplings and anisotropies  by renormalized one  [$J_n \rightarrow J'_n$  and $\Delta_n \rightarrow \Delta'_n$ in the Eq.~(\ref{eq2})],
 and results
\begin{eqnarray}
{\cal H}_{n}^{(i,\alpha;\; j,\beta)}= J'_{n}
(
\sigma^{x}_{i,\alpha}\sigma^{x}_{j,\beta}
+
\sigma^{y}_{i,\alpha}\sigma^{y}_{j,\beta}
)
+
\Delta'_{n}
\sigma^{z}_{i,\alpha}\sigma^{z}_{j,\beta}
.
\end{eqnarray}
%
These renormalized coupling constants are functions of the original ones which are given by the following equations
%
\begin{eqnarray}
\label{RG-equations}
\bl
J'_{1}&=
\varepsilon_{1}
J_{1};
\;\;\;\;
\;\
\;\;
\;
\Delta'_{1}=\varepsilon_{1}
\Big[
\frac{\delta_{1  }}{2}
\Big]^{2}
\Delta_{1},
\\
J'_{d}&=\varepsilon_{1}
J_{d};
\;\;\;\;\;
\;\;\;\;
\Delta'_{d}=
\Big[
\frac{\delta_{1  }}{2}
\Big]^{2}
\varepsilon_{1}
\Delta_{d},
\\
J'_{0}&=
-\Big[
\frac{2}{\delta_{1  }}
\Big]
\varepsilon_{1}
 J_{d}
+
\Big[
\frac{2\delta_{1  }^{2}+1}{\delta_{1  }^{2}}
\Big]
\varepsilon_{1}
J_{0},
 \\
\Delta'_{0}&=
\Big[
\frac{2-\delta_{1  }^{2}}{2}
\Big]
\varepsilon_{1}
 \Delta_{d}
+\Big(\frac{3\delta_{1  }^{4} -4(\delta_{1  }^{2}-1)}{4\delta_{1  }^{2}} \Big)
\varepsilon_{1}
\Delta_{0},
\el
\end{eqnarray}
%
where
$
\varepsilon_{1}=
\Big[
\frac{2\delta_{1  }}{2+\delta_{1  }^{2}}
\Big]^{2},
$
and
\bea
\delta_{n  }=\frac{1}{2J_{n}}
\Big(
\Delta_{n}+\sqrt{\Delta_{n}^{2}+8J_{n}^2}
\Big);
\;\;\;\; (n=0,1,d).
\eea
The rescaled  QRG equations
can be  obtained by dividing  the above equations by  the factor of $\varepsilon_{1}$. Furthermore, the stable and unstable fixed points of the rescaled equations are evaluated  by  solving the following equations
\begin{eqnarray}
\bl
J'_{n}=J_n=J^{*}_{n};\;\;\;\;
\Delta'_{n}=\Delta_{n}=\Delta^{*}_{n},
\el
\eea
 For simplicity and without loss of generality, we restrict our analysis to the case of isotropic interaction in leg and diagonal couplings, $\Delta_{1}=J_{1},~\Delta_{d}=J_{d}\;\; (\delta_{1  }=2)$.
  This restrictive case will suffice to show the interesting feature of the system.
In the isotropic case the renormalized coupling constants [rescaled of Eq.~(\ref{RG-equations})] are reduced the following form
%
\begin{eqnarray}
\bl
&
J'_{1}=\Delta'_{1}=\Delta_{1}=J_{1},
\;\;\;\;\;\;
J'_{d}=\Delta'_{d}=\Delta_{d}=J_{d},
\\
&
J'_{0}=-J_{d}+ \frac{9}{4} J_{0};
\;\;\;\;\;\;\;\;\;
\;\;
\Delta'_{0}=\!-\!J_{d}+\frac{9}{4}\Delta_{0}.\;\;\;\;\;\;\;\;\;\;\;\;
\el
\label{eq3}
\end{eqnarray}
%
%
 Thus, the QRG equations express that the isotropic exchange interaction of leg and diagonal coupling are preserved under QRG, $\Delta'_{1}=J_{1},~\Delta'_{d}=J'_{d}$, and  surprisingly diagonal coupling,  $J_{d}$, dose not flow under QRG transformations.
 Since the diagonal coupling induces frustrated magnetic orders,
  therefore the universality class of the model is unaffected in the presence of the frustrating interaction.
Moreover, for each value of $J_{d}$, there is a critical point $\Delta_{0}^{c}=J_{0}^{c}=0.8J_{2}$ above which plaquette coupling $J_{0}$ and corresponding anisotropy $\Delta_{0}$ go to infinity ($J_{0}\rightarrow +\infty,~\Delta_{0}\rightarrow +\infty$),
 while for $J_{0}<J_{0}^{c}$ and $\Delta_{0}<\Delta_{0}^{c}$ the plaquette coupling and plaquette anisotropy decrease gradually and their sign change from positive (antiferromagnetic) to negative (ferromagnetic) after a few steps, and run finally to infinity ($J_{0}\rightarrow -\infty,~\Delta_{0}-\infty$).

%
\begin{figure}[t]
\includegraphics[width=1.0\columnwidth]{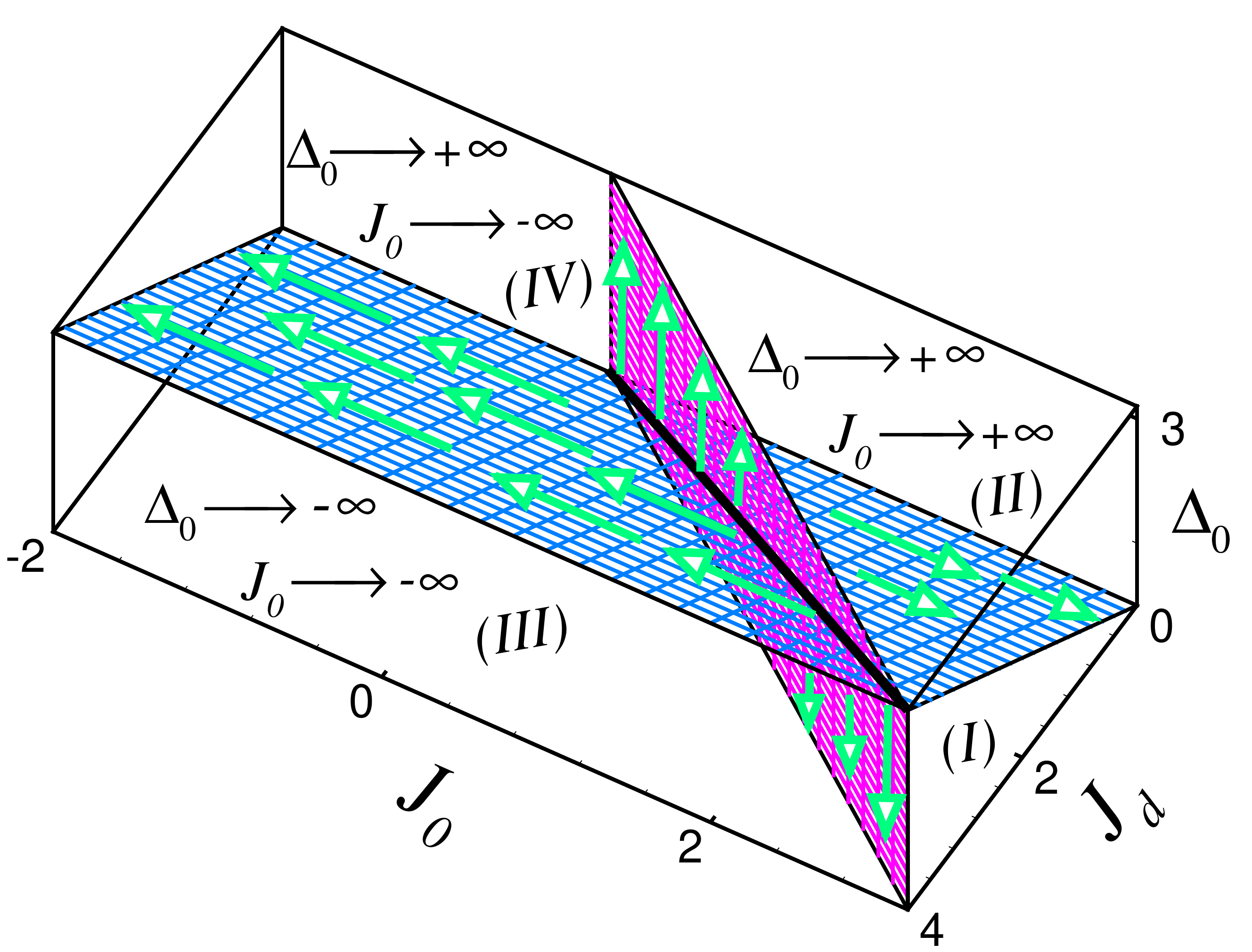}
\caption{(Color online)  Three dimensional phase
diagram of four-legs spin tube in a absence of magnetic
field. The blue and purple planes are critical surfaces
where separate the phase diagram of model into four distinct
region. The flow of couplings are different in each region.
Arrows show the running of couplings under QRG.
The black solid line corresponds to the critical line
where the blue critical surface meets with the purple critical
surface.
We set
$J_{1}=1.$
}
\label{fig3}
\end{figure}
%

The QRG flow shows that the model has four stable fixed points (lines) located at ($J_{d},~J_{0}\rightarrow \mp\infty, ~\Delta_{0}\rightarrow \mp\infty$) while ($J_{d},~J_{0}=0.8J_{d},~\Delta_{0}$) and ($J_{d},~J_{0},~\Delta_{0}=0.8J_{d}$) stand for the unstable fixed points which specify the critical surfaces of the model (Fig. \ref{fig3}).
The significant result of our calculations occurs when $J_{0}\longrightarrow0$, in which the spin nanotube decouples to four weakly interacting XXZ chains and can be analyzed by means of bosonization and conformal field theory~\cite{Cabra1998, Cabra1997}.
The QRG equations show that, in the presence of very small diagonal coupling, if we start with $J_{0}=0$, the ferromagnetic plaquette interaction ($J_{0}<0$) is generated under QRG transformation and runs to infinity. The generation of plaquette interaction under QRG originates from the presence of diagonal coupling.

We have linearized the QRG flow at the critical line ($J_{d},~J_{0}^{c}=0.8J_{d},~\Delta_{0}^{c}=0.8J_{d}$) (black thick solid line in Fig.~\ref{fig3}) and found two relevant and one marginal directions. The eigenvalues of the matrix of linearized flow are $\lambda_{1}=\lambda_{2}=\frac{9}{4}$ and $\lambda_{3}=1$.
The corresponding eigenvectors in the $|J_{d},J_{0},\Delta_{0}\rangle$ coordinates are $|\lambda_{1}\rangle=|0,1,0\rangle$,
$|\lambda_{2}\rangle=|0,0,1\rangle$ and $|\lambda_{3}\rangle=|\frac{5}{4},1,1\rangle$. The relevant directions ($|\lambda_{1}\rangle,~|\lambda_{2}\rangle$)
 show the flow direction of plaquette coupling $J_{0}$ and plaquette anisotropy $\Delta_{0}$,
 respectively. The marginal direction ($|\lambda_{3}\rangle$) corresponds to the tangent line of the critical line where a critical surface ($J_{d},~J_{0}^{c}=0.8J_{d},~\Delta_{0}$) meets with a critical surface ($J_{d},~J_{0},~\Delta_{0}=0.8J_{d}$) (Fig.~\ref{fig3}).
We have also calculated the critical exponents at the critical line ($J_{d},~J_{0}^{c}=0.8J_{d},~\Delta_{0}^{c}=0.8J_{2}$)~\cite{Delgado19962}.
In this respect, we have obtained the dynamical exponent and the diverging exponent of the correlation length.
The dynamical exponent is given by
\bea
z =
\frac{
\ln
\Big(
\frac{J'_{1}}{J_{1}}\Big)
}{
\ln(n_{B})
}
=2,
\eea
where $n_{B}=3$ is the number of sites in each block. The correlation length diverges as
\bea
\xi\sim|J_{0}-J^{c}_{0}|^{-\nu}\sim|\Delta_{0}-\Delta^{c}_{0}|^{-\nu}
\eea
 with exponent
$\nu=1.35$, which is expressed by
\bea
\nu=
\frac{\ln(n_{B})}{
\ln
\Big(
\frac{dJ'_{0}}{dJ_{0}}
\Big)
}
=
\frac{
\ln(n_{B})
}{
\ln
\Big(
\frac{d\Delta'_{0}}{d\Delta_{0}}
\Big)
}.
\eea
\\

It is remarkable to note that, in the absence of a magnetic field, the frustrating NNN leg interaction, $J_{2}$, is generated automatically under QRG transformation by adding the second order corrections~\cite{Jafari2006, Jafari2007}. For small NNN leg interaction ($J_{2}<J_{1}$) the QRG equations show running of $J_{2}$ to zero except at the isotropic point ($\Delta_{1}=J_{1}$) where $J_{2}$ runs to the tri-critical point $J_{2}=0.155J_{1}$ ~\cite{Jafari2006, Jafari2007}.  To reduce complexity, in this paper the second order correction is not  considered.

%
\subsubsection{In the presence of the magnetic field: $h\neq0$}
In the presence of an external magnetic field, the QRG analysis can be done in a manner analogous to the work done in the zero magnetic field case. The only difference is that, due to the level crossing which occurs at
%
\begin{eqnarray}
h_{1}=\frac{1}{8}
\Big(
3\Delta_{1}+\sqrt{\Delta_{1}^{2}+8J_{1}^2}
\Big)
=\frac{1}{4}
(
\Delta_{1}+ \delta_{1}J_{1}
),
\eea
%
 for the eigenstates of the block Hamiltonian, the projection operator $P_{0}$ can be different depending on the coupling constants (see Appendix~\ref{App_C}). The first order effective Hamiltonian for $h<h_{1}$, is similar to the original one, Eq. (\ref{eq1}), and the normalized couplings, apart from a renormalized magnetic field, is exactly the same as the zero magnetic field case, Eq. (\ref{eq3}), and the renormalized magnetic field is given by
 $$h'=
 \Big(\frac{2+\delta_{1  }^{2}}{2\delta_{1  }}
 \Big)^{2}
 h.$$
The process of renormalization of Hamiltonian Eq. (\ref{eq1}) for $h>h_{1}$, to the first order corrections, leads to the similar Hamiltonian with different coupling constants given in the Appendix~\ref{App_C}.
In the presence of an external magnetic field,  for simplicity, we restrict ourself to the case of isotropic exchange interactions in intra-leg ($\Delta_{1}=J_{1}$) and diagonal ($\Delta_{d}=J_{d}$) couplings.
The running of couplings under QRG transformation for $h<h_{1}$ shows that the magnetic field increases gradually and goes beyond the level crossing point ($h_{1}$) after a few steps, which means both regions $h<h_{1}$ and $h>h_{1}$ are unique phases.
Thus, it is sufficient to study the QRG-flows of the system just for $h>h_{1}$ to obtain the fixed points, critical points and the ground state phase of the system.\\

For $h>h_{1}$ the QRG-flows show running of leg coupling $J_{1}$ to zero, which represents the renormalization of the energy scale, and the initial isotropic case ($\Delta_{d}=J_{d}$) are not preserved under QRG transformation.
Moreover, starting with any initial values of  $\Delta_{1}$ and $\Delta_{d}$, the intra-leg and diagonal anisotropies run to zero, while the plaquette interaction and corresponding anisotropy go toward infinity for any initial values of $J_{0}$ and $\Delta_{0}$. It is necessary to mention that the diagonal interaction $J_{d}$ does not flow under QRG even in the presence of a magnetic field.
In summary, QRG equations express that the spin tube prepared in the strong leg-coupling limit goes to the strong plaquette coupling limit which has been considered in the following section.
%
\begin{figure}
\includegraphics[width=1.0\columnwidth]{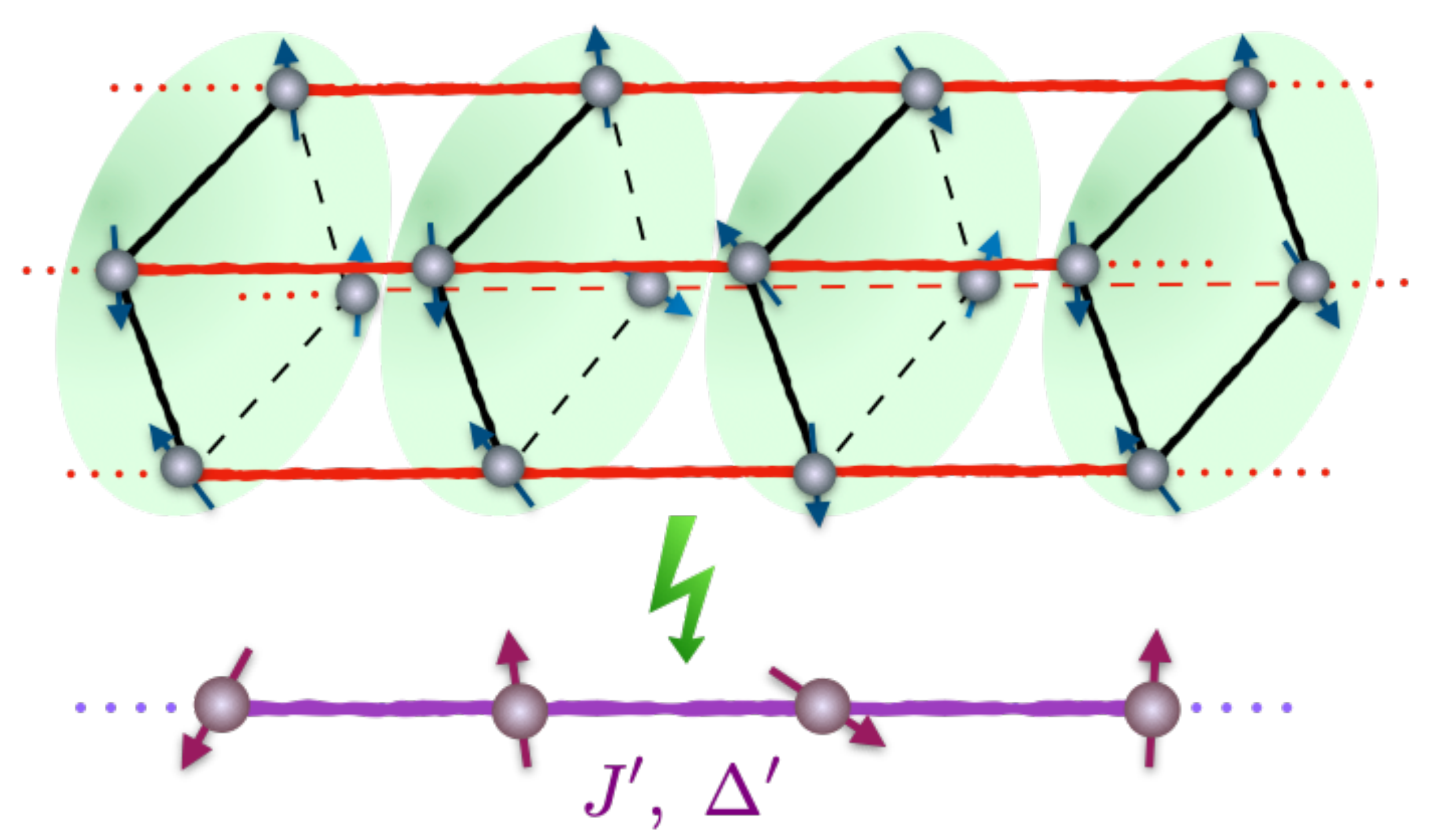}
\caption{(Color online)  A schematic  decomposition of
four-legs spin tube into plaquettes blocks where each
plaquette replaced by an effective single site under
renormalization process.}
\label{fig4}
\end{figure}
%

\subsection{The strong plaquette coupling (Weakly interacting plaquettes): $J_{0}/J_{1,d}\rightarrow\infty$ }
\label{Se-weak-limit}
In  the strong plaquette coupling limit where we have
$J_{0}/J_{1,d}\rightarrow\infty$, the plaquettes are  almost decoupled, and the inter-plaquette couplings $J_{1,d}$ can be dealt with perturbatively.
To apply the QRG scheme to the model in the strong plaquette coupling limit, we consider  the original  Hamiltonian,  Eq.~(\ref{eq1}), 
and  split the spin tube into blocks that each contains independent plaquette (see Fig.~\ref{fig4}).
In that matter, the Hilbert space of each plaquette has sixteen states including  two spin-$0$ singlets, nine spin-$1$ triplets and five spin-$2$ quintuplets~\cite{Gomez}.
The four lowest eigenvalues of the plaquette Hamiltonian labeled by $e_{\alpha=0,1,2,3}$ and their corresponding eigenstates are given in the Appendix~\ref{App_D}.
Because of a possible energy level crossing  between these eigenstates, the projection operator, $P_{0}$, can be different depending on the coupling constants.
We classify the regions  corresponding two lowest eigenvalues to construct their projection operators, and we discuss the phase diagram in terms of the following five different regions:
%
\subsubsection{\bf Region I: $h<\Delta_{0}-J_{0}$\hfill~}
In this region we consider  $e_{0}$ as a ground state and $e_{2}$ as a first excited state, and to the first order
corrections the effective Hamiltonian  leads to the   exactly solvable  1D   transverse field Ising model 
%
\begin{eqnarray}
\label{eq5}
{\cal H}^{eff}=-\sum_{i=1}^{N}\Big[ J'  \sigma^{x}_{i}\sigma^{x}_{i+1}+h' \sigma^{z}_{i}\Big],
\end{eqnarray}
%
where
\begin{equation}
\bl
J' =\frac{\delta_{0  }^{2}}{(2+\delta_{0  }^{2})} (\Delta_{d}-\Delta_{1});\;\;\;
h' =\delta_{0  }J_{0} -\Delta_{0} .
\el
\end{equation}
The 1D Ising model in a transverse field
is exactly solvable by the Jordan-Wigner transformation~\cite{Lieb} and the RSRG~\cite{Delgado19962}.
For simplicity we consider isotropic interaction on the plaquette $\Delta_{0}=J_{0}$. Then,
the renormalized coupling and transverse field reduce to
$J'=(\Delta_{d}-\Delta_{1})/3$, and
$h' =J_{0}$.
Phase transition between the paramagnetic and antiferromagnetic/ferromagnetic phases takes place at
$h' =J'_{}$ under which the system is ferromagnet ($\Delta_{1}<\Delta_{d}$) or antiferromagnet ($\Delta_{1}>\Delta_{d}$) while the system
enters the paramagnetic phase above the critical point $h' >J'_{}$.
It is remarkable that,  by assuming equal anisotropy ratios for the leg and diagonal interactions $\Delta_{1}=\Delta_{d}$
the system is always in the paramagnetic phase where spins aligned along the direction of the external magnetic field.
%
\subsubsection{\bf  Region II: $\Delta_{0}-J_{0}<h< (\delta_{0  }-1)J_{0}$\hfill~ }
In this region, we have
$e_{0}$ as a ground state and $e_{1}$ as a first excited state.
%
This leads  the effective Hamiltonian  to the well known 1D XXZ model in the presence of an external magnetic field (Appendix~\ref{App_E}), which can be solved exactly by the Bethe ansatz method~\cite{Cloizeaux,Yang}
%
%
\begin{equation}
\label{eq6}
{\cal H}^{eff}
\!=\!
\sum_{i=1}^{N}\Big[\frac{J' }{4}(\sigma^{x}_{i}\sigma^{x}_{i+1}+ \sigma^{y}_{i}\sigma^{y}_{i+1})
+\frac{\Delta' }{4}\sigma^{z}_{i}\sigma^{z}_{i+1}-\frac{h' }{2}\sigma^{z}_{i}\Big].
\end{equation}
%
%
Here  couplings of renormalized Hamiltonian are given by
\begin{eqnarray}
\label{couplings_II}
\label{eq7}
\bl
J' &=\frac{(\delta_{0  }+2)^{2}}{(4+2\delta_{0  }^{2})}  (J_{1}-J_{d});
\;\;\;\;
\;\;\;\;
\Delta' =\frac{1}{4}(\Delta_{1}+\Delta_{d}),\;\;\;\;
\;\;\;\;
\;\;
\\
h' &=-  h+(\delta_{0  }-1)J_{0}+\Delta' .
\el
\end{eqnarray}


%
\subsubsection{\bf Region III: $(\delta_{0  }-1)J_{0}
<h<
\frac{1}{2} (\Delta_0+\delta_{0  }J_{0})$\hfill~
}
This region is defined by the situation of
$e_{1}$ as a ground state and $e_{0}$ as a first excited state,
one can show that  the first-order effective Hamiltonian is the same as former  case, Eq.~(\ref{eq6}),  with the negative field, and the
couplings are defined as  before, Eq.~(\ref{couplings_II}).
%


%
\subsubsection{\bf Region IV: $\frac{1}{2} (\Delta_0+\delta_{0  }J_{0})<h<\Delta_{0}+J_{0}$ \hfill~}
%
For the  field in this interval,
$e_{1}$ is a ground state and $e_{3}$ is a first excited state, and
the effective Hamiltonian is also similar to the case of regions II with different coupling constants defined by
%
\begin{equation}
\label{couplings_IV}
\bl
J' &=(J_{1}-J_{d});
\;\;\;\;\;\;
 ~~\Delta' =\frac{1}{4}(\Delta_{1}+\Delta_{d}),
\\
h' &=-h+J_{0}+\Delta_{0}+3\Delta' .
\el
\end{equation}
%

\subsubsection{\bf Region V: $h>\Delta_{0}+J_{0}$ \hfill~} 
For the field which fulfills $h>\Delta_{0}+J_{0}$,  $e_{3}$ is the ground state and $e_{1}$ is the first excited state. 
Therefore,  the effective Hamiltonian up to the first order is the same as the region IV with the magnetic field in opposite direction, and the coupling constants are the same as Eq.~(\ref{couplings_IV}).
%

%
\begin{figure*}[t]
\vspace{-0.25cm}
\centerline{
\includegraphics[width=0.975\linewidth]{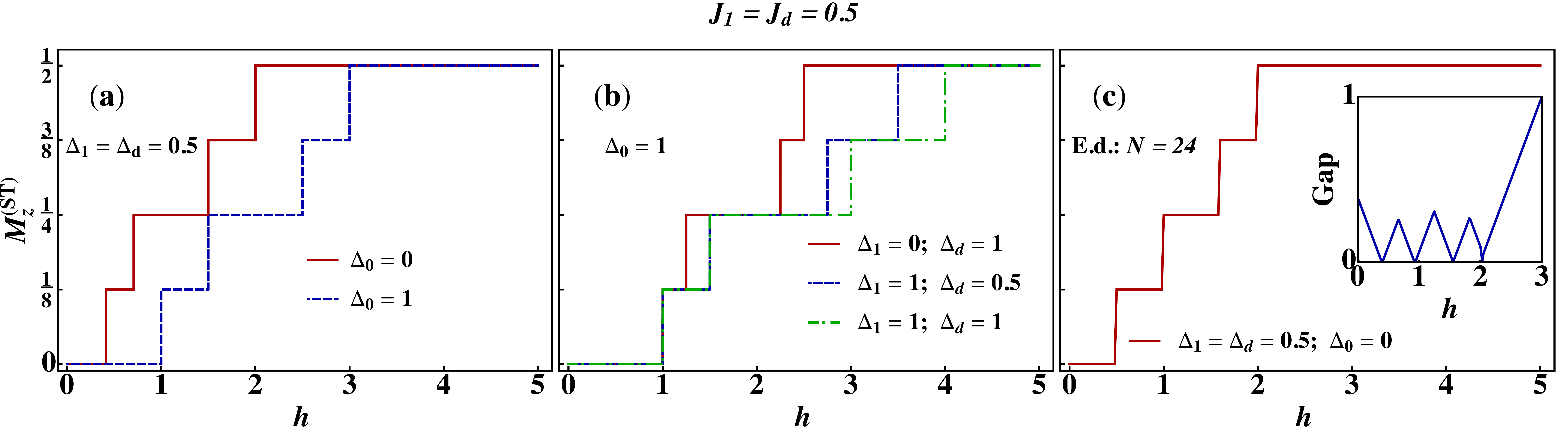}
}
\vspace{-0.3cm}
\caption{ (Color online) Magnetization (per site) of FAFST versus the magnetic
field $h$ obtained by QRG transformation on the maximum frustration line $J_{1}=J_{d}$, with couplings and anisotropic parameters:
(a) $\Delta_1=\Delta_d=0.5J_{0}$ and two cases of $\Delta_0=0$ and  $\Delta_0=J_{0}$, and
(b)   $\Delta_0=J_{0}$ and three cases of
$\Delta_1=0;\;\Delta_d=J_{0}$,
$\Delta_1=J_{0};\;\Delta_d=0.5J_{0}$, and
$\Delta_1=J_{0};\;\Delta_d=J_{0}$.
(c)
Same quantity obtained by the numerical Lanczos results on finite spin-1/2 nanotube systems with size $N=24$,
 for different values of the exchanges as $\Delta_0=0$,  and $J_1=\Delta_1= J_d=\Delta_d=0.5J_{0}$. The inset shows the corresponding energy~gap.
 Here and in the following figures,  we set $J_{0}=1$.
}
\label{fig5}
\end{figure*}

%
\subsection{Phase Transition}
As shown, the renormalized Hamiltonian, in the strong plaquette coupling limit, is different than  the original
one, FAFST, to find the recursion relation.
However,  the effective Hamiltonian are exactly solvable~\cite{Cloizeaux,Yang,Langari1998}
and it enables us to predict distinct features of the spin tube in the strong plaquette coupling limit.
To prevent the complexity, we restrict our study to the case $ \Delta_{0}=J_{0}$ and $h\geq0$.
In such a case, our analysis does not cover the region I and we only consider the regions II-V,
where the expected FAFST models in the presence of the magnetic field are mapped to the well-known 1D spin $1/2$ exactly solvable models.

\subsubsection{First order phase transition: $J_{1}= J_{d}$}
In the case of the equal  inter-plaquette couplings $J_1=J_d$, the frustration is maximum, and  the effective model reduces to
the well-known 1D     spin-1/2 Ising model in a longitudinal magnetic field,
%
\begin{eqnarray}
\label{eq11}
{\cal H}^{eff}=\frac{1}{4}\sum_{i=1}^{N}\Big[\Delta' \sigma^{z}_{i}\sigma^{z}_{i+1}\pm2h' \sigma^{z}_{i}\Big],
\end{eqnarray}
The ground state properties of this model has been
investigated using the RSRG method~\cite{Langari2004}. This model shows a first order transition from a
classical antiferromagnetic ordered phase to the saturated ferromagnetic phase at $\Delta' =h' $.
Depends on the values of the anisotropy parameter $\Delta' $ and the magnetic field $h' $,
the effective Hamiltonian reveals two magnetization (per site) plateaus $M_{z}^{eff}=0$ and $M_{z}^{eff}=\pm1/2$
correspond to the antiferromagnetic and the ferromagnetic phases, respectively.
Consequently, the first order phase transition points of the FAFST are given by
%
\begin{equation}
\bl
h_{c_{1}}&=h_{c_{2}}=\delta_{0  }-J_{0},
\\
h_{c_{3}}&=h_{c_{4}}=\delta_{0  }-J_{0}+\frac{1}{2}(\Delta_{1}+\Delta_{d}),
\\
\label{eq12}
h_{c_{5}}&
\!=
\!
h_{c_{6}}
\!
=J_{0}+\Delta_{0}+\frac{1}{2}(\Delta_{1}+\Delta_{d}),
\\
h_{c_{7}}&
\!
=
\!
h_{c_{8}}
\!
=J_{0}+\Delta_{0}+\Delta_{1}+\Delta_{d}.
\el
\end{equation}
%
Additionally, the magnetization in the FAFST are connected to the magnetization plateaus
in the effective Hamiltonian using the renormalization equation (see the Appendix~\ref{App_D}).

For  $h<\frac{1}{2} (\Delta_0+\delta_{0  } J_{0})$ 
the plateaus at $M^{eff}_{z}=0$, $M^{eff}_{z}=\pm1/2$ in the curve of magnetization (per site) versus  $h' $ in the effective model translate
into plateaus at $M^{ST}_{z}=1/8$, $M^{ST}_{z}=0$, and $M^{ST}_{z}=1/4$ in the curve of magnetization (per site) versus $h$ in the FAFST.
There are also renormalization equation for $h>\frac{1}{2} (\Delta_0+\delta_{0  }J_{0})$ for linking the magnetization plateaus in the effective
Hamiltonian to the magnetization curve in the FAFST (see  Appendix~\ref{App_D}).
 In such case, plateaus at $M^{eff}_{z}=0$, $M^{eff}_{z}=\pm1/2$ in the magnetization curve of the effective model turn into $M^{ST}_{z}=1/4$, $M^{ST}_{z}=3/8$ and $M^{ST}_{z}=1/2$ in the magnetization curve of the FAFST model.
The magnetization curves of FAFST along the maximum frustration line $(J_{1}=J_{d})$ have been shown in
Figs.~\ref{fig5}(a~and~b) based on the numerical RSRG results.
To examine the anisotropy effects, the magnetization curves of FAFST have been plotted versus the magnetic field $h$ for different values of anisotropies.
Notice that in Fig.~\ref{fig5}(a) the magnetization plateaus has been depicted versus $h$ for isotropic case
 $\Delta_{0}=J_{0}, J_{1}=\Delta_{1}, J_{d}=\Delta_{d}$ (dashed-blue curve) which shows quantitatively excellent agreement with numerical density matrix renormalization group results~\cite{Gomez}.
This result indicates  that the RSRG is a good approach to study the critical behavior of FAFST in the thermodynamic limit.

To examine the anisotropy effects, the magnetization curves of FAFST have been plotted versus the magnetic field $h$ in Fig.~\ref{fig5} for different values of anisotropies.
 As seen, the location of critical points and the width of the  magnetization plateaus are controlled by the anisotropies
according to Eq. (\ref{eq12}). It is to be noted that, the renormalized subspace specified by the singlet $(e_{0})$ and triplet $(e_{1})$
states is separate at the level crossing point $h_{l}=\frac{1}{2} (\Delta_0+\delta_{0  }J_{0})$ from the renormalized subspace defined by the triplet $(e_{1})$ and quintuplet $(e_{3})$ states. Therefore, for the cases that
$h_{c_{3}}~(=h_{c_{4}})$
 is greater than $h_{l}$, the point $h_{c_{3}} $ is not the critical point and the level crossing point would be a first order phase transition point (Figs.~\ref{fig5}(a)-(b)).
From Fig.~\ref{fig5}(b) one can clearly see that the width of $M^{(ST)}_{z}=1/8$ and $M^{(ST)}_{z}=3/8$ plateaus reduced
by decreasing the inter-plaquette anisotropies ($\Delta_{1}+\Delta_{d}$).

To accomplishment of  our study, using the numerical Lanczos method we have studied the effect of the external magnetic field on the ground state magnetic phase diagram of the mentioned FAFST model.
 In Fig.~\ref{fig5}(c), we have presented our numerical results.
 In this figure on top of the magnetization, in the inset  we  plot   the energy gap  as a function of the magnetic field for a tube size $N=24$ and different values of the exchanges according to the $\Delta_0=0$, $J_1=\Delta_1=0.5 J_{0}$ and $J_d=\Delta_d=0.5 J_{0}$.
 As is seen, in the absence of the magnetic field the FAFST model is gapped. By increasing the magnetic field, the energy gap decreases linearly and vanishes at the first critical field.
 By more increasing the magnetic field, the energy gap will be closed in other three critical magnetic fields,  
  independent of the system size.
 After the fourth critical field, the gap opens again and for a sufficiently large field becomes proportional to the magnetic field which is known as the indication of the ferromagnetic phase.
 On the other hand, the magnetization is zero in the absence of the magnetic field at zero temperature. By increasing the magnetic field, besides the zero and saturation plateaus, three magnetization plateau at
$M =1/8 , M = 2/8, M =3/8$ are observed.  We have to mention that the critical fields estimated by the numerical Lanczos method are in complete agreement with our analytical results presented in figures Fig.~\ref{fig5}(a).
\\

\begin{figure}[t]
\centerline{
\includegraphics[width=0.96\linewidth]{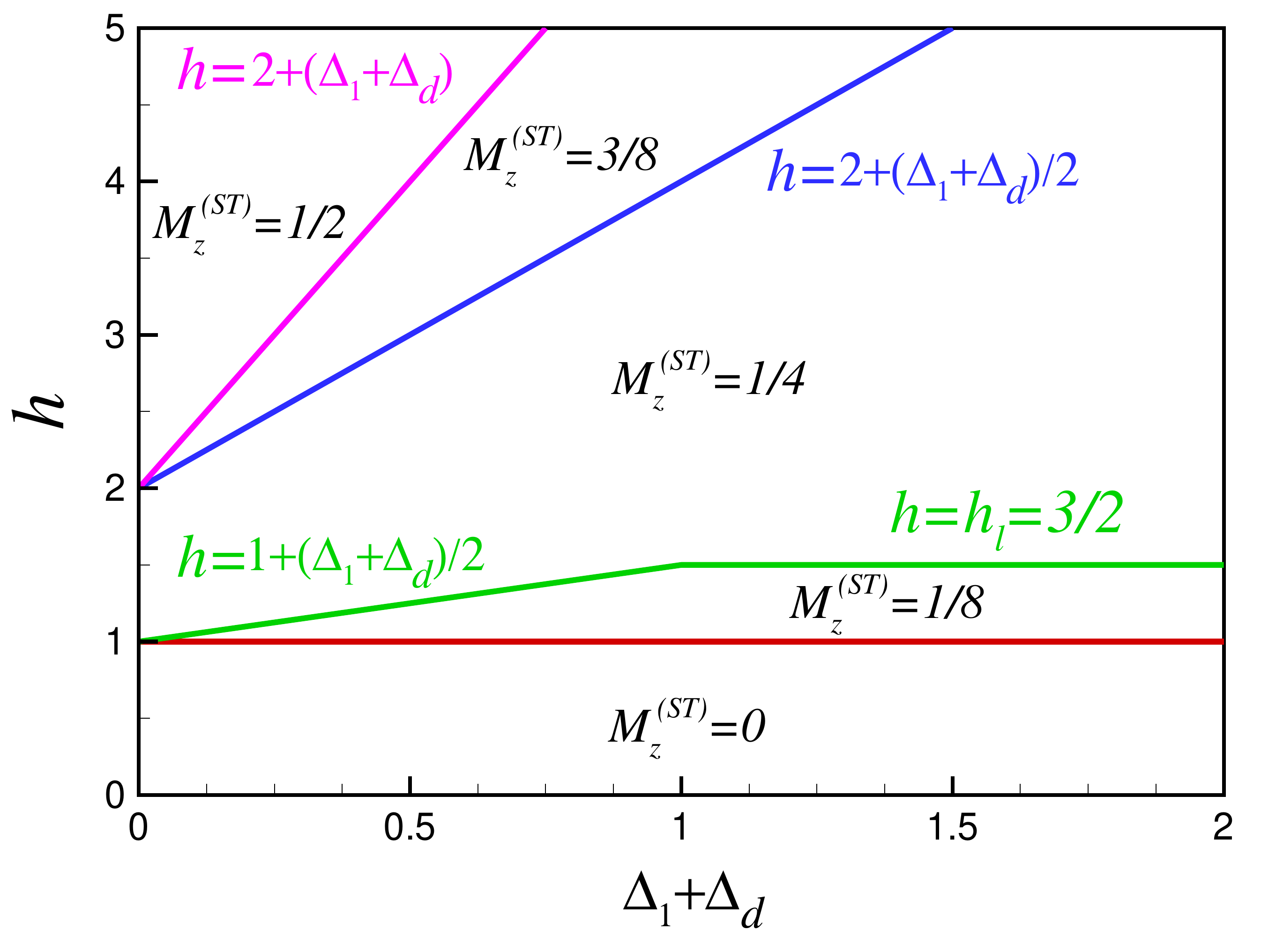}
}
\vspace{-0.3cm}
\caption{ (Color online)
The phase diagram: $h$ vs $(\Delta_{1}+\Delta_{d})$ along the
line $J_{1}=J_{2}$ obtained by QRG transformation for $\Delta_{0}=J_{0}=1$.
}
\label{fig6}
\end{figure}

%
\begin{figure*}[t!]
\vspace{-0.25cm}
\centerline{
\includegraphics[width=0.975\linewidth]{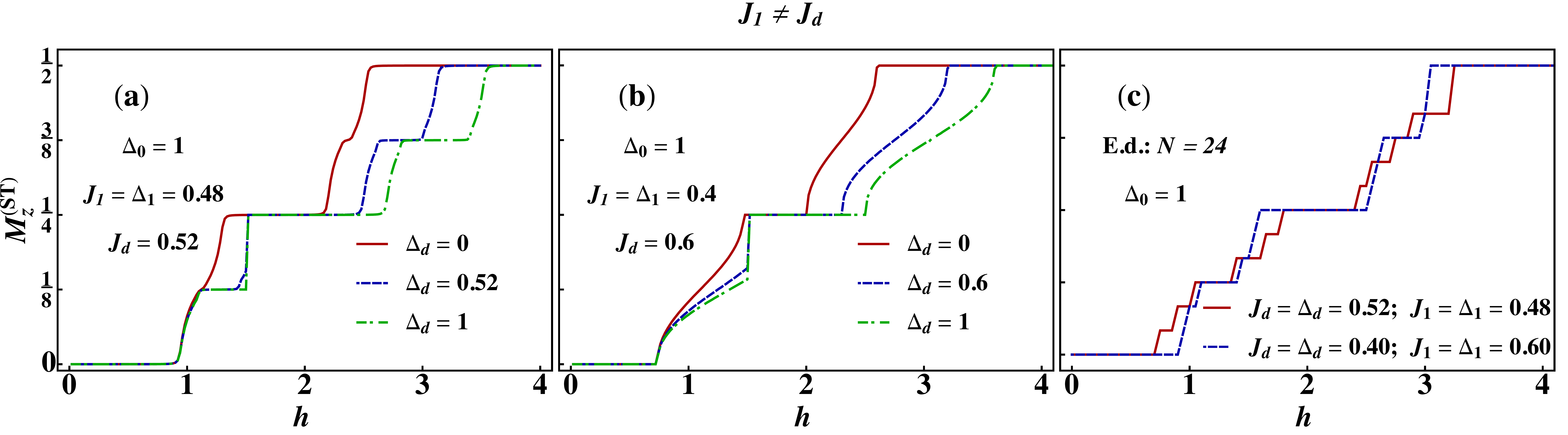}
}
\vspace{-0.35cm}
\caption{ (Color online) Magnetization (per site) of FAFST versus the magnetic
field $h$ obtained by QRG transformation for $J_{1}\neq J_{d}$ cases,
with couplings and anisotropic parameters:
(a)  $\Delta_0=J_{0};\; J_1=\Delta_1=0.48J_{0};\; J_d=0.52J_{0}$ and three cases of    $\Delta_d=(0,0.52,1)J_{0}$, and
(b)  $\Delta_0=J_{0};\; J_1=\Delta_1=0.4J_{0};\; J_d=0.6J_{0}$ and three cases of    $\Delta_d=(0,0.6,1)J_{0}$.
(c)
Numerical Lanczos results of the same quantity on finite spin-1/2 nanotube systems with the system  size $N=24$.
We set $\Delta_0=J_{0}$, and the rest of  exchanges are considered for two set of parameters as
$J_1=\Delta_1=0.48J_{0};\; J_d=\Delta_d=0.52J_{0}$ (solid-red) and
$J_1=\Delta_1=0.4J_{0}; \; J_d=\Delta_d=0.6J_{0}$ (dashed-blue).
}
\label{fig7}
\end{figure*}

The magnetic phases of FAFST along maximum frustration line $J_{1}=J_{d}$ has been shown versus $(\Delta_{1}+\Delta_{d})$ and $h$ in Fig.~\ref{fig6} for $\Delta_{0}=J_{0}$, based on the RSRG approach.
As it can be observed, $M^{(ST)}_{z}=3/8$ and $M^{(ST)}_{z}=1/2$, plateaus width linearly increase with frustrating anisotropy $(\Delta_{1}+\Delta_{d})$, while width of $M^{(ST)}_{z}=1/8$ plateau initially increases linearly with $(\Delta_{1}+\Delta_{d})$ and then at $h_{l}$ point reaches to the constant value.

It would be worth mentioning that although at the isotropic point: $\Delta_{0}= J_{0}, J_{1}=\Delta_{1}, J_{d}=\Delta_{d}$, the critical points of FAFST,  Eq.~(\ref{eq12}), reduces to the critical points expression obtained by low energy effective method~\cite{Gomez}, but the magnetic phase obtained by QRG method is not the same as that of obtained by the low energy effective method~\cite{Gomez}.
This discrepancy originates from the presence of level crossing point $h_{l}$, where the system shows first order first transition. The low energy effective method is incapable of capturing the effect of this level crossing point even away from the maximum frustration line $J_{1}=J_{d}$.

\subsubsection{Second order phase transition: $J_{1}\neq J_{d}$}

As we mentioned previously, in the case where inter-plaquette couplings are not equal $J_{1}\neq J_{d}$, the FAFST Hamiltonian maps to the effective Hamiltonian, the 1D spin 1/2 XXZ chain in the presence of the longitudinal magnetic field.
This model is exactly solvable by means the Bethe ansatz method. Moreover, the properties of the XXZ model in the presence of a magnetic field has been studied using the QRG method~\cite{Langari1998}. In this subsection we study the effective Hamiltonian, by combining a Jordan-Wigner transformation~\cite{Lieb} with a mean-field approximation~\cite{Caux} (see Appendix \ref{App_E}). Then by using the renormalization equations which connects the magnetization of the effective model to that of FAFST, we can obtain the magnetization of the FAFST.
Therefore, it is useful to briefly review the main features of the 1D XXZ chain in the presence of the magnetic field.\\
In absence of a magnetic field $h'=0$, for $\Delta'= J_{0}$, the Hamitonian is $SU(2)$ symmetry invariant, but for $\Delta'\neq  J_{0}$ the $SU(2)$ symmetry breaks down to the $U(1)$ rotational symmetry around the $z$-axis. It is known that for planar anisotropy $- J_{0}\leq\Delta'\leq  J_{0}$ the model is  not supporting any kind of long range order where the correlations decay algebraic and the ground state is gapless~\cite{Haldane}, so called Luttinger liquid phase. Enhancing the amount of anisotropy stabilizes the spin alignment. For $\Delta'> J_{0}$, the symmetry of the ground state  is reduced to $Z_{2}$ and the ground state is the gapped N\`{e}el ordered state which is in the universality class of 1D antiferromagnetic Ising chain. Indeed, the third term in the Hamiltonian Eq. (\ref{eq6}) causes N\`{e}el ordering in the system while the first two terms in the Hamiltonian extend the quantum fluctuations in the system and result in the corruption of the ordering. Furthermore, for $\Delta<- J_{0}$ the ground state is the gapfull ferromagnetic state.

In presence of a magnetic field $h'\neq0$, there are two critical lines $h_{c1}$ and $h_{c2}$
restricting Luttinger liquid phase between ferromagnetic and N\`{e}el phases which are given by following equations~\cite{Yang}
%
\begin{eqnarray}
\bl
&
h_{c1}=J'\sinh g\sum_{n=-\infty}^{+\infty}\frac{(-1)^{n}}{\cosh(ng)};
\\
&
 h_{c2} =\pm \Big(1+\frac{\Delta'}{J_{0}} \Big) |J'|,
\el
\end{eqnarray}
%
with $ g=\cosh^{-1}(\frac{\Delta'}{J_{0}}).$
For $\Delta'>J_{0}$ and small magnetic fields $h'<h_{c}$ the ground state is still the N\`{e}el ordered state.
This state exhibits a gap in the excitation spectrum whose value at $h'=0$ corresponds to $h_{c1}$.
In particular, it is exponentially small close to the Heisenberg point at $\Delta'=J_{0}$, which is characteristic for a Kosterlitz-Thouless transition~\cite{Kosterlitz1973, Kosterlitz1974}. The Luttinger-liquid state exists for $|\Delta'|<J_{0}$, $h'<h_{c2}$ and $\Delta'>J_{0}$, $h_{c1}<h'<h_{c2}$.
Finally, phase transition between Luttinger-liquid and ferromagnetic occurs at $h=h_{c_2}$ under which the ground state is the ferromagnetically polarized state along the $z$-direction.\\

To study the effect of anisotropy, the magnetization of FAFST is plotted versus the magnetic field in Fig.~\ref{fig7}, for different values of anisotropies.
It can be clearly seen that, $M^{(ST)}_{z}=1/8$ and $M^{(ST)}_{z}=3/8$ plateaus width enhances (reduces) by increasing (decreasing) the inter-plaquette anisotropies ($\Delta_{1}, \Delta_{d}$), and the first order phase transition point at $h_{l}$ fades out gradually by decreasing $\Delta_{1}$ and $\Delta_{d}$.
%
As seen, for a small deviations from the maximum frustrated line $J_{1}=0.48J_{0}, J_{d}=0.52J_{0}$, width of $M^{(ST)}_{z}=1/8$ and $M^{(ST)}_{z}=3/8$ plateaus reduce, and jumps between plateaus change to smooth curves which is feature of Luttinger liquid phases.
As represented in Fig.~\ref{fig7}(b), $M^{(ST)}_{z}=1/8$ and $M^{(ST)}_{z}=3/8$ magnetization plateaus are not present for $J_{1}=0.4J_{0}, J_{d}=0.6J_{0}$. Away from the maximum frustrated line, the magnetization shows only a gapless Luttinger liquid phase, which means the system consists of decoupled spin-1/2 chains.
In other words, the presence of $M^{(ST)}_{z}=1/8$ and $M^{(ST)}_{z}=3/8$ plateaus is very sensitive to frustration.

Again we have implemented our numerical Lanczos algorithm on the mentioned FAFST model. In Fig.~\ref{fig7}(c), we have presented our numerical results for different values of the anisotropy parameters.
In this figure, the  magnetization for a size  $N=24$ is plotted as a function of the magnetic field for two set of anisotropy parameters according to  $\Delta_0=J_{0}$; $J_1=\Delta_1=0.48J_{0}$,  $J_d=\Delta_d=0.52J_{0}$ and  $\Delta_0=J_{0}$; $J_1=\Delta_1=0.4J_{0}$,  $J_d=\Delta_d=0.6J_{0}$.
As is seen in this figure, the place and width of magnetic plateaues are in complete agreement with the analytical results presented in Fig.~\ref{fig7}(b).
One should note that observed oscillations of the magnetization in the Fig.~\ref{fig7}(c) are arised from the level crossings in the finite size systems.\\

Inspecting  the effect of anisotropy,   clearly shows  the $M^{(ST)}_{z}=1/8$ and $M^{(ST)}_{z}=3/8$ plateaus width enhances (reduces) by increasing (decreasing) the inter-plaquette anisotropies ($\Delta_{1}, \Delta_{d}$), and the first order phase transition point at $h_{l}$ fades out gradually by decreasing $\Delta_{1}$ and $\Delta_{d}$.
To summarize, for small deviation from maximum frustrating line, the one-eight and third-eight magnetization plateaus width, which are sensitive to frustration, can be controlled by the intra-plaquette anisotropies. Away from the maximum frustrating line, where the  one-eighth and three-eighth magnetization plateaus are absent, the intra-plaquette anisotropies can affect the width of one-quarter magnetization plateau.

\section{Summary}
We survey a geometrically frustrated anisotropic four-leg spin tube 
in the absence/presence of the magnetic field using the quantum real Space  Renormalization group.
We show that in the limit of the strong leg coupling the diagonal frustrating interaction does not fellow under renormalization transformation.
Moreover, 
 in the limit of the strong leg coupling, the spin tube goes to the strong plaquette coupling limit, i.e., weakly interacting plaquettes.
Our study indicates that  in the weakly interacting plaquettes, FAFST  maps onto the 1D spin-1/2 XXZ model  under renormalization transformation.
For the case that the leg and frustrating couplings are the same (maximum frustrating line), the FAFST Hamiltonian reveals only first order quantum phase transitions at zero temperature.
In such a case,    fractional magnetization plateaus at zero, one-quarter, one-half and three-quarter of the saturation magnetization are exhibited.
Also, the magnetization plateaus at one-quarter and three-quarter of the saturation magnetization show the highest sensitivity to frustration and washed out away from the maximum frustrating line.
Comparing the QRG to the density matrix renormalization group results guaranteed that the real space renormalization group method is a remarkable approach to study the critical behavior of FAFST in the thermodynamic limit.
We have also calculated the classical phase diagram by considering the spin structure as a spiral~\cite{Arlego2011}.
The result show that $M^{ST}_{z}=0$ and $M^{ST}_{z}=1/2$ can be captured by classical spin while, absence of $M^{ST}_{z}=1/8$, $M^{ST}_{z}=1/4$, and $M^{ST}_{z}=3/8$, shows that they are originated from quantum effect (frustration).

At the end,  attention should be paid to the importance of the exploration of quantum correlation, such as entanglement and quantum discord in the FAFST model. They can be easily done in
the thermodynamics limit by applying the real Space  Renormalization group method~\cite{Usman2017, Jafari2013, Langari2012, Jafari2010, Jafari2008}. In particular, studying the four-leg spin tube frustrated by next-nearest-neighbor interaction on the leg,  is an interesting topic that clearly deserves future investigations, which are not
 considered in the current work.

%
\section*{ACKNOWLEDGMENTS}
A.\!\;A. acknowledges financial support through National Research Foundation (NRF) funded by the Ministry of Science of Korea (Grants  No. 2017R1D1A1B03033465, \& No. 2019R1H1A2039733),
  and by the National Foundation of Korea  funded by the Ministry of Science, ICT and Future Planning (No. 2016K1A4A4A01922028).
%
\section*{Appendices}
\appendix
\section{Real Space  Renormalization Group (RSRG)}
\label{App_A}
When dealing with zero temperature properties of many-body systems with a large number
of strongly correlated degrees of freedom, one can consider real space  renormalization Group (RSRG)
as a one of the possible and powerful  methods~\cite{Delgado19961, Delgado19962, Langari2004, Langari2012}.
Where, its application   on lattice systems implies the construction of a new smaller system corresponding to the original one with new (renormalized) interactions between the degrees of freedom~\cite{SAJafari2018, SAJafari2017, Jafari2017}.
One of the tasks of QRG is to obtain the recursion relation, which define the transformation of old couplings into new ones. In  the Kadanoffs representation, analyzing
such recursion relation  determines  qualitatively the structure of the phase diagram; 
approximately locates the critical/fixed points   and obtains the critical exponents~\cite{Delgado19962, Langari2004, Jafari2007}.
This  method divides a lattice  into disconnected blocks of $n_{B}$ sites each where the Hamiltonian is exactly diagonalized. This partition of the lattice
into blocks induces a decomposition of the Hamiltonian ${\cal H}$  into an intrablock Hamiltonian ${\cal H}_{B}$ and a interblock Hamiltonian ${\cal H}^{BB}$,
where the block Hamiltonian ${\cal H}^{B}$ is a sum of commuting terms, ${\cal H}^{B}=\sum_{I=1}^{n/n_{B}}h^{B}_{I}$, each acting on different ($I$th) blocks of chain.
Each block is treated independently to build the projection operator $P_{0}$ onto the
lower energy subspace. The projection of the Hamiltonian is mapped to an effective Hamiltonian ${\cal H}^{eff}$
 acts on the renormalized subspace.   
Thus, in perturbative approach, the effective Hamiltonian up to first order corrections is given by~\cite{Delgado19962}
%
\begin{eqnarray}
{\cal H}^{eff}={\cal H}_{0}^{eff}+{\cal H}_{1}^{eff},
\eea
with
\bea
\nonumber
{\cal H}_{0}^{eff}=P_{0}{\cal H}^{B}P_{0};\;\;\;\;
{\cal H}_{1}^{eff}=P_{0}{\cal H}^{BB}P_{0}.
\end{eqnarray}

\section{The intrea-block and intre-block Hamiltonians of three sites, its eigenvectors and eigenvalues in the absence of magnetic field\label{Apendix A}}
\label{App_B}
The inter-block ${\cal H}^{BB}$ and intra-block ${\cal H}^{B}$  Hamiltonians for
the three sites decomposition are
%
\begin{equation}
\bl
h_{I}^{B}=
\frac{1}{4}
&
\sum_{\alpha=1}^{4}
\Bigg[
J_{1}
\Big(
\sigma^{x}_{1,\alpha,I}\sigma^{x}_{2,\alpha,I}
+ \sigma^{y}_{1,\alpha,I}\sigma^{y}_{2,\alpha,I}
+\sigma^{x}_{2,\alpha,I}\sigma^{x}_{3,\alpha,I}
\\&
+ \sigma^{y}_{2,n,I}\sigma^{y}_{3,\alpha,I}
\Big)
+\Delta_{1}
\Big(
\sigma^{z}_{1,\alpha,I}\sigma^{z}_{2,\alpha,I}
+\sigma^{z}_{2,\alpha,I}\sigma^{z}_{3,\alpha,I}
\Big)
\Bigg];
\el
\end{equation}
%
%
\begin{equation}
\label{HBB-nonmag}
\bl
&H^{BB}=\frac{1}{4}\sum_{I=1}^{N/3}\sum_{\alpha=1}^{4}
\Bigg[
\\
&
J_{1}
\Big(
\sigma_{3,\alpha,I}^{x}\sigma_{1,\alpha,I+1}^{x}+
\sigma_{3,\alpha,I}^{y}\sigma_{1,\alpha,I+1}^{y}
\Big)
+\Delta_{1}\sigma_{3,\alpha,I}^{z}\sigma_{1,\alpha,I+1}^{z}
\\
&
+J_{0}
\Big(
\sigma_{1,\alpha,I}^{x}\sigma_{1,\alpha+1,I}^{x}
+\sigma_{2,\alpha,I}^{x}\sigma_{2,\alpha+1,I}^{x}
+\sigma_{3,\alpha,I}^{x}\sigma_{3,\alpha+1,I}^{x}
\\
&
+\sigma_{1,\alpha,I}^{y}\sigma_{1,\alpha+1,I}^{y}
+\sigma_{2,\alpha,I}^{y}\sigma_{2,\alpha+1,I}^{y}
+\sigma_{3,\alpha,I}^{y}\sigma_{3,\alpha+1,I}^{y}
\Big)
\\
&
+\Delta_{0}
\Big(
\sigma_{1,\alpha,I}^{z}\sigma_{1,\alpha+1,I}^{z}
+\sigma_{2,\alpha,I}^{z}\sigma_{2,\alpha+1,I}^{z}
+\sigma_{3,\alpha,I}^{z}\sigma_{3,\alpha+1,I}^{z}
\Big)
\\
&+J_{d}
\Big(
\sigma_{3,\alpha,I}^{x}\sigma_{1,\alpha+1,I+1}^{x}+
\sigma_{3,\alpha,I}^{y}\sigma_{1,\alpha+,I+1}^{y}
\Big)
\\
&
+\Delta_{d}\sigma_{3,\alpha,I}^{z}\sigma_{1,\alpha+1,I+1}^{z}
\Bigg],
\el
\end{equation}
%
where $\sigma_{j, \alpha,I}^{\mu}$ refers to the $\mu$-component of
the Pauli matrix at site $j$ of the block labeled by $I$ with itra-plaquettes label $\alpha$. The exact
treatment of this Hamiltonian leads to four distinct eigenvalues
which are doubly degenerate.
The ground, first, second and third
excited state energies have the following expressions in terms of
the coupling constants:
\begin{equation}
\bl
&
e_{0}=-\frac{1}{2}\delta_{1  }J_{1}
;\;\;
e_{1}=\frac{1}{2}\delta'_{1  }J_{1}
;\;\;
e_{2}=0;\;\;
%
e_{3}=\frac{ \Delta_{1}}{2},
\el
\end{equation}
with corresponding eigenfunctions
%
\begin{equation}
\bl
&|\psi_{0}\rangle=
\frac{1}{\sqrt{2+\delta_{1  }^{2}}}(|\uparrow\uparrow\downarrow\rangle
-
\delta_{1  }|\uparrow\downarrow\uparrow\rangle+|\downarrow\uparrow\uparrow\rangle),
\\
&|\psi_{0}'\rangle=
\frac{1}{\sqrt{2+\delta_{1  }^{2}}}(|\uparrow\downarrow\downarrow\rangle
-
\delta_{1  }|\downarrow\uparrow\downarrow\rangle+|\downarrow\downarrow\uparrow\rangle),\\
&|\psi_{1}\rangle=
\frac{1}{\sqrt{2+\delta_{1  }^{'2}}}(|\uparrow\uparrow\downarrow\rangle+
\delta'_{1  }|\uparrow\downarrow\uparrow\rangle+|\downarrow\uparrow\uparrow\rangle),
\\
&|\psi_{1}'\rangle=
\frac{1}{\sqrt{2+\delta_{1  }^{'2}}}(|\uparrow\downarrow\downarrow\rangle+
\delta'_{1  }|\downarrow\uparrow\downarrow\rangle+|\downarrow\downarrow\uparrow\rangle),
\\
&|\psi_{2}\rangle=
\frac{1}{\sqrt{2}}(|\downarrow\downarrow\uparrow\rangle-
|\uparrow\downarrow\downarrow\rangle),
\\
&|\psi_{2}'\rangle=
\frac{1}{\sqrt{2}}(|\uparrow\uparrow\downarrow\rangle-
|\downarrow\uparrow\uparrow\rangle),
\\
&
|\psi_{3}\rangle=|\uparrow\uparrow\uparrow\rangle,
\\
&
|\psi_{3}'\rangle=|\downarrow\downarrow\downarrow\rangle,
\el
\end{equation}
%
%
where
\begin{equation}
\delta'_{n  }=\delta_{n  }-
\frac{\Delta_{n}}{J_{n}}=\frac{1}{2J_{n}}
\Big[\sqrt{
\Delta_{n}^{2}+8J_{n}^2
}
-\Delta_{n}
\Big],
\end{equation}
and  we consider
$|\uparrow\rangle$ and $|\downarrow\rangle$ as the eigenstates of $\sigma^{z}$.

\section{The intrea-block and intre-block Hamiltonians of three sites, its
eigenvectors and eigenvalues in the presence of magnetic field\label{Apendix B}}
\label{App_C}
In the present of field the inter-block  Hamiltonian 
for
the three sites decomposition are
%
\begin{equation}
\bl
h_{I}^{B}=
&
\sum_{\alpha=1}^{4}\frac{1}{4}\Bigg[
J_{1}\Big(
\sigma^{x}_{1,\alpha,I}\sigma^{x}_{2,\alpha,I}+ \sigma^{y}_{1,\alpha,I}\sigma^{y}_{2,\alpha,I}
+\sigma^{x}_{2,\alpha,I}\sigma^{x}_{3,\alpha,I}
\\
&
+ \sigma^{y}_{2,\alpha,I}\sigma^{y}_{3,\alpha,I}
\Big)
+\Delta_{1}
\Big(
\sigma^{z}_{1,\alpha,I}\sigma^{z}_{2,\alpha,I}
+\sigma^{z}_{2,\alpha,I}\sigma^{z}_{3,\alpha,I}
\Big)
\\
&
-2h(\sigma^{z}_{1,\alpha,I}+\sigma^{z}_{2,\alpha,I}+\sigma^{z}_{3,\alpha,I})\Bigg],
\el
\end{equation}
%
%
and intra-block 
Hamiltonian ${\cal H}^{BB}$ is defined in similar way as Eq.~(\ref{HBB-nonmag}).
The ground, first, second and third
excited state energies have the following expressions in terms of
the coupling constants:
\bea\no
\bl
&e_{0/1}=-\frac{1}{2}
(
\pm h+ \delta_{1  } 
);
\;\;\;
e_{2}=\frac{1}{4}(2\Delta_{1}-6h);
\;\;\;
e_{3}=-\frac{1}{2}h;\;\;\;
\\
&e_{4/5}=\frac{1}{2}
\Big
[\mp h +\delta'_{1  } 
\Big];
\;\;\;\;
e_{6}=\frac{1}{2}h;
\;\;\;\;
e_{7}=\frac{1}{2}(\Delta_{1}+6h),
\el
\eea
with following eigenfunctions
\bea
\bl
&|\psi_{0}\rangle=\frac{1}{\sqrt{2+\delta_{1  }^{2}}}(|\uparrow\uparrow\downarrow\rangle
-
\delta_{1  }|\uparrow\downarrow\uparrow\rangle+|\downarrow\uparrow\uparrow\rangle),
\\
&|\psi_{1}\rangle=\frac{1}{\sqrt{2+\delta_{1  }^{2}}}(|\uparrow\downarrow\downarrow\rangle
-
\delta_{1  }|\downarrow\uparrow\downarrow\rangle+|\downarrow\downarrow\uparrow\rangle),
\\
&
|\psi_{2}\rangle=|\uparrow\uparrow\uparrow\rangle,
\\
&
|\psi_{3}\rangle=\frac{1}{\sqrt{2}}(|\downarrow\uparrow\uparrow\rangle
-|\uparrow\uparrow\downarrow\rangle),
\\
&|\psi_{4}\rangle=\frac{1}{\sqrt{2+\delta_{1  }^{'2}}}(|\uparrow\uparrow\downarrow\rangle+
\delta'_{1  }|\uparrow\downarrow\uparrow\rangle+|\downarrow\uparrow\uparrow\rangle),
\\
&|\psi_{5}\rangle=\frac{1}{\sqrt{2+\delta_{1  }^{'2}}}(|\uparrow\downarrow\downarrow\rangle+
\delta'_{1  }|\downarrow\uparrow\downarrow\rangle+|\downarrow\downarrow\uparrow\rangle),
\;\;\;\;
\\
&
|\psi_{6}\rangle=\frac{1}{\sqrt{2}}(|\downarrow\downarrow\uparrow\rangle-
|\uparrow\downarrow\downarrow\rangle),
\\
&
|\psi_{7}\rangle=|\downarrow\downarrow\downarrow\rangle.
\el
\eea
The   renormalized and rescaled  coupling constants for $h>h_{1}$ are defined by
%
\bea
\bl
J'_{1}=&
J_{1};
\;\;\;\;\;\;
\;\;
\;\;
\Delta'_{1}=(\frac{1}{2+\delta_{1  }^{2}}) \Delta_{1},
\\
J'_{d}=&
J_{d};
\;\;\;
\;\;\
\;
\;\;\;
\Delta'_{d}=(\frac{1}{2+\delta_{1  }^{2}}) \Delta_{d},
\\
J'_{0}=&
-2 \delta_{1  } J_{d} +(2+\delta_{1  }^{2}) J_{0} ,
\\
\Delta'_{0}=&
(\frac{2\delta_{1  }^{2}}{2+\delta_{1  }^{2}}) \Delta_{d}
+(\frac{2+\delta_{1  }^{4}}{2+\delta_{1  }^{2}}) \Delta_{0},
\\
h'=
&( 2+\delta_{1  }^{2} )(- h+ \Delta_{1}+\delta_{1  } J_{1})
+2(\frac{1+\delta_{1  }^{2}}{2+\delta_{1  }^{2}}) \Delta_{1}
\;\;\;\;
\\
&+2(\frac{5+3\delta_{1  }^{2}+2\delta_{1  }^{4}}{2+\delta_{1  }^{2}})\Delta_{d}
+2(\frac{2+4\delta_{1  }^{2}}{2+\delta_{1  }^{2}})\Delta_{0}
.
\el
\eea
Here the rescaling factor  is $\varepsilon'_{1}=(\frac{1}{2+\delta_{1  }^{2}})$.
%

\section{The plaquette Hamiltonian, four lowest eigenvalues of plaquette Hamiltonian and their corresponding eigenstates in weak leg couplings\label{Apendix C}}
\label{App_D}
In the strong plaquette coupling (Weakly interacting plaquettes: $J_{0}/J_{1,d}\rightarrow\infty$) one can write the  inter-block and intra-block Hamiltonians for
the plaquette decomposition as
%
%
\begin{equation}
\bl\no
h_{I}^{B}=
\frac{1}{4}
&
\Bigg[
J_{0}
\Big(
\sigma^{x}_{1,I}\sigma^{x}_{2,I}
+\sigma^{x}_{2,I}\sigma^{x}_{3,I}
+\sigma^{x}_{3,I}\sigma^{x}_{4,I}
+\sigma^{x}_{4,I}\sigma^{x}_{1,I}
\\&
+\sigma^{y}_{1,I}\sigma^{y}_{2,I}
+\sigma^{y}_{2,I}\sigma^{y}_{3,I}
+\sigma^{y}_{3,I}\sigma^{y}_{4,I}
+\sigma^{y}_{4,I}\sigma^{y}_{1,I}
\Big)
\\&
+\Delta_{0}
\Big(\sigma^{z}_{1,I}\sigma^{z}_{2,I}
+\sigma^{z}_{2,I}\sigma^{z}_{3,I}+\sigma^{z}_{3,I}\sigma^{z}_{4,I}
+\sigma^{z}_{4,I}\sigma^{z}_{1,I}
\Big)
\\&
-2h
\Big(\sigma^{z}_{1,I}+\sigma^{z}_{2,I}+\sigma^{z}_{3,I}+\sigma^{z}_{4,I}
\Big)\Bigg],
\el
\\
\end{equation}
%
%
and
%
\begin{equation}
\bl\no
{\cal H}^{BB}&
\!
=
\frac{1}{4}\sum_{I=1}^{N}
\Bigg[
\Delta_{d}
\Big(\sigma_{1,I}^{z}\sigma_{2,I+1}^{z}
+\sigma_{2,I}^{z}\sigma_{3,I+1}^{z}
+\sigma_{3,I}^{z}\sigma_{4,I+1}^{z}
\Big)
\\&
+\!
\sum_{\alpha=1}^{4}
\Big[\!
J_{1}
\Big(
\sigma_{\alpha,I}^{x}\sigma_{\alpha,I+1}^{x}
+\sigma_{\alpha,I}^{y}\sigma_{\alpha,I+1}^{y}
\Big)
\!+\!
\Delta_{1}\sigma_{\alpha,I}^{z}\sigma_{\alpha,I+1}^{z}
\Big]
\\&
+J_{d}
\Big(
\sigma_{1,I}^{x}\sigma_{2,I+1}^{x}
+\sigma_{2,I}^{x}\sigma_{3,I+1}^{x}
+\sigma_{3,I}^{x}\sigma_{4,I+1}^{x}
\\&
\hspace{1cm}+
\sigma_{1,I}^{y}\sigma_{2,I+1}^{y}
+\sigma_{2,I}^{y}\sigma_{3,I+1}^{y}
+\sigma_{3,I}^{y}\sigma_{4,I+1}^{y}
\Big)
\Bigg],
\el
\\
\end{equation}
here $\sigma_{\alpha,I}^{\mu}$ refers to the $\mu$-component of
the Pauli matrix at the block labeled by $I$ with itra-plaquettes label $\alpha$. The  eigenstates are obtained  by
%
\begin{equation}
\bl
|\psi_{0}\rangle=
&\frac{1}{\sqrt{4+2\delta_{0  }^{2}}}
(|\uparrow\uparrow\downarrow\downarrow\rangle
-
\delta_{0  }|\uparrow\downarrow\uparrow\downarrow\rangle+|\downarrow\uparrow\uparrow\downarrow\rangle
\\
&
-\delta_{0  }|\downarrow\uparrow\downarrow\uparrow\rangle+|\downarrow\downarrow\uparrow\uparrow\rangle+|\uparrow\downarrow\downarrow\uparrow\rangle),
\;\;\;
\\
|\psi_{1}\rangle=&
\frac{1}{2}(|\uparrow\uparrow\downarrow\uparrow\rangle
-|\uparrow\uparrow\uparrow\downarrow\rangle+|\downarrow\uparrow\uparrow\uparrow\rangle-|\uparrow\downarrow\uparrow\uparrow\rangle),
\\
|\psi_{2}\rangle=&
\frac{1}{\sqrt{2}}(|\downarrow\uparrow\downarrow\uparrow\rangle-|\uparrow\downarrow\uparrow\downarrow\rangle)
\\
|\psi_{3}\rangle=&
|\uparrow\uparrow\uparrow\uparrow\rangle,
\el
\end{equation}
 and corresponding eigenvalues are given by
%
\begin{equation}
\bl\no
e_{0}= -\delta_{0  } J_{0};\;\; 
e_{1}=- (J_{0}+h) ;\;\;
e_{2}=- \Delta_{0};\;\;   e_{3}=(\Delta_{0}-2h) .
\el
\end{equation}
The magnetization (per site) in the effective Hamiltonian $M^{eff}_{z}$ linked to the magnetization (per site) of the spin tube $M^{ST}_{z}$ through the renormalization transformation
of the $\sigma^{z}$ component of the Pauli matrices. The $\sigma^{z}$ in the effective Hilbert space
has the following transformations ($\alpha=1,2,3,4$) for each  region:
\\
%
\begin{itemize}
  \item  II: $P_{0}^{I}\sigma_{\alpha,I}^{z}P_{0}^{I}=\frac{1}{4}(1-\sigma_{I}^{z}); ~~M^{ST}_{z}=\frac{1}{8}(1-2M^{eff}_{z})$,

   \item III:  $P_{0}^{I}\sigma_{\alpha,I}^{z}P_{0}^{I}=\frac{1}{4}(1+\sigma_{I}^{z}); ~ M^{ST}_{z}=\frac{1}{8}(1+2M^{eff}_{z})$,

    \item IV:  $P_{0}^{I}\sigma_{\alpha,I}^{z}P_{0}^{I}=\frac{1}{4}(3-\sigma_{I}^{z}); ~ M^{ST}_{z}=\frac{1}{8}(3-2M^{eff}_{z})$,

    \item V:  $P_{0}^{I}\sigma_{\alpha,I}^{z}P_{0}^{I}=\frac{1}{4}(3+\sigma_{I}^{z});~ ~M^{ST}_{z}=\frac{1}{8}(3+2M^{eff}_{z})$.
\end{itemize}
\vspace{0.5cm}

\section{Mean field approximation\label{Apendix D}}
\label{App_E}
One of the analytical approaches to study the renormalized 1D spin-1/2 XXZ Hamiltonian (Eq.~(\ref{eq6})) is the fermionization technique. In this respect, by applying the Jordan-Wigner transformation~\cite{Lieb} the effective XXZ Hamiltonian maps onto a system of interacting spinless fermions\cite{Dmitriev02}:
%
\begin{equation}
\label{eqAD1}
\bl
{\cal H}=
\sum_{i}
&
\Big[\frac{J' }{2}(c^{\dag}_{i} c_{i+1}+c^{\dag}_{i+1} c_{i})+\Delta' c^{\dag}_{i} c_{i}c^{\dag}_{i+1} c_{i+1}
\Big]
 \\
&-(h'+\Delta')\sum_{i}c^{\dag}_{i} c_{i} .
\el
\end{equation}
%
It is known that the fermion interaction term can be decomposed by following relevant order parameters which are related to spin-spin correlation functions as\cite{Dmitriev02}
%
\begin{eqnarray}
\label{eqAD2}
\gamma_{1}&=
\langle c^{\dag}_{i}c_{i}
\rangle
;\;\;\;~\gamma_{2}&=
\langle
c^{\dag}_{i}c_{i+1}
\rangle,
\end{eqnarray}
%
Using $c^{\dag}_{i} = \frac{1}{\sqrt{N}} \sum ^{N} _{n=1} e^{ikn} c^{\dag}_{k}$,  the Fourier transformation to momentum space is performed and then the diagonalized Hamiltonian is obtained as
%
\begin{eqnarray}
\label{eqAD3}
{\cal H}=
\sum_{k} \varepsilon(k) c^{\dag}_{k} c_{k},
\end{eqnarray}
%
where the energy spectrum is
%
\begin{eqnarray}
\varepsilon(k)=(J-2\gamma_2 \Delta ) \cos (k)-(h+\Delta-2 \gamma_1\Delta ).
\end{eqnarray}
%
Easily, one can show that the Fermi points are given by
%
\begin{eqnarray}
\pm k_F=\pm \arccos\!\Bigg[\frac{h+\Delta-2\gamma_1 \Delta }{J-2\gamma_2 \Delta }
\Bigg].
\label{fermi-points}
\end{eqnarray}
%
One should note that the following equations should be satisfied self-consistently
%
\begin{eqnarray}
\gamma_1=1-\frac{k_F}{\pi},~\gamma_2=-\frac{\sin(k_F)}{\pi},
\label{self-1}
\end{eqnarray}
%
Finally, the magnetization is obtained as
%
\begin{eqnarray}
\bl
M= \left\{ \begin{array}{ll}
        \frac{1}{2} &   \hspace{1.25cm} h>h_{c}\\
        -\frac{1}{2}&  \hspace{1.25cm}   h<-h_{c}\\
      \frac{1}{2}-\frac{k_F}{\pi}&   \hspace{0.75cm}  -h_c<h<h_{c}
        \end{array}
         \right.
         \el
\end{eqnarray}
%
\bibliography{Ref}

\end{document}